\documentclass[aps,prl,twocolumn,showpacs,preprintnumbers]{revtex4}
\usepackage{amssymb}
\usepackage{multirow}
\usepackage[colorlinks,linkcolor=blue,citecolor=blue,dvipdfm]{hyperref}
\usepackage{graphicx}
\usepackage{dcolumn}
\usepackage{bm}
\usepackage{color,soul}

\begin{document}

\title{Superconducting Resonance and paring symmetry in electron-doped cuprates}
\date{\today}

\begin{abstract}
The magnetic excitations in the superconducting electron-doped cuprates are studied in the framework of spin-density-wave description. The superconducting resonance is a natural product of the superconductivity due to the opening of d-wave gap. Its resonance energy exhibits well linear scaling with superconducting gap as $E_{res}/2\Delta\sim 0.6$, quantitatively consisting with the experimental discovery. This ratio is insensitive to the selected parameters, manifesting its universality. Another lower-energy peak below resonance energy is predicted when the hole pocket emerges due to suppression of spin-density wave. We further verify that the ratio of linear scales is intimately related to the pairing symmetry. Distinct ratio can be found with respective pairing symmetry. In comparison with the inelastic neutron scattering data,  the monotonic d-wave superconductivity is the most likely candidate in the electron-doped cuprates. Furthermore, we proposed a new method to check the pairing symmetry by the inelastic neutron scattering measurements.
\end{abstract}

\pacs{74.25.Ha, 74.72.Ek, 74.72.Kf, 74.20.Mn}
\author{H. Y. Zhang$^1$, Y. Zhou$^1$, C. D. Gong$^2,^1$, and H. Q. Lin$^3$ }
\affiliation{$^1$National Laboratory of Solid State Microstructure, Department of
Physics, Nanjing University, Nanjing 210093, China \\
$^2$Center for Statistical and Theoretical Condensed Matter Physics,
Zhejiang Normal University, Jinhua 321004, China \\
$^3$Department of Physics and the Institute of Theoretical Physics, Chinese
University of Hong Kong, Hong Kong, China}
\maketitle
\date{\today }

High-temperature superconductivity is one the most challenging topics in condense matter physics. It arises from the charge carrier doping into their insulating antiferromagnetic parent compounds\cite{Armitage-RMP}. Due to the proximity of antiferromagnetism (AFM) and superconductivity (SC), it is generally believed that there exists intrinsic link between them. The spin fluctuations is often proposed to be glue of pairing in the cuprates. Understanding the nature of the spin fluctuation and its relation with superconductivity are essential for the mechanism of superconductivity. The inelastic neutron scattering(INS) provides direct way to investigate the spin dynamics.

The electron-doped cuprates provide unique opportunity to study the competition between the AFM and SC due to its special replacement and oxygen annealing process.\cite{Zhao-NPHY11}. So far, extensive INS experiments had been performed in electron-type cuprates\cite{Zhao-NPHY11,Fujita-PRL08,Yamada-PRL03,Zhao-PRL07,Wilson-Nature06,Wilson-PNAS07,Yu-NPHY09,Yu-PRB10} (For more details, see Ref.\cite{FujitaRev}). Except the well known commensurate magnetic excitations in the normal state\cite{Yamada-PRL03}, another universal feature in the superconducting state is further discovered. There exists a superconducting resonance, an unusual spin-triplet collective mode at $\mathbf{Q}=(\pi, \pi)$. For example, the resonance energy $E_{res}$ is about $9.5 meV$, and $11 meV$, in the optimal doping $NCCO$%
\cite{Zhao-PRL07}, and $PLCCO$\cite{Wilson-Nature06}, respectively. The
relative $E_{res}$ and its associated condensation energy can be
suppressed through applying external magnetic field\cite{Wilson-PNAS07} or
oxygen annealing process\cite{Zhao-NPHY11}. However, the ratio of $%
E_{res}/k_{B}T_{c}$ is almost unchanged with fixed value about $5.8$%
\cite{Wilson-Nature06,Zhao-PRL07}. However, Yu \emph{et al.} argued that the universal scaling is $E_{res}/\Delta$ rather than $E_{res}/T_{c}$.\cite{Yu-NPHY09}. Furthermore, the single superconducting resonance may be constituted by two separated sub-peaks\cite{Yu-PRB10}. Interestingly, the linear scale is fairly universal as can also been found in the hole-doped cuprates, heavy-fermion compounds, and some Fe-pnictides (For more details, see Ref.~\cite{Yu-NPHY09}).

Based on the kinetic energy driven superconducting mechanism, a dome shaped doping dependent resonance energy is proposed in electron-doped cuprates. However, the intensity at given energy in the SC state is almost three orders of magnitude lager than that of the normal state\cite{Feng}, inconsistent with the nearly unchanged feature in experiments. Ismer \emph{et al.} showed the resonance can be regarded as a overdamped collective mode located near the particle-hole continuum\cite{Ismer-PRL07}. Their results indicated that the resonance energy is sensitive to the selected parameter. Therefore, the linear scaling $E_{res}/\Delta$ or $E_{res}/k_{B}T_{c}$ is not expected in this framework. Furthermore, those theories based on the single band description\cite{LJX-PRB03,Kruger-PRB07} can not account the properties of spin dynamics, for example, the commensurate magnetic excitation, as we argued in the previous paper\cite{HY}. To our knowledge, the possible linear scaling of superconducting resonance and its intrinsic relation with the SC have not well established.

In this paper, the superconducting resonance is studied in details in the electron-doped cuprates within the framework of spin-density-wave (SDW) description. The superconducting resonance exhibits well linear scaling with superconducting gap as $E_{res}/2\Delta\sim 0.6$. This ratio is universal, and is insensitive to the selected parameters. Another superconducting sub-resonance develops when the SDW is suppressed. We further verify that the linear scales and its ratio are the intrinsic nature of monotonic d-wave pairing by inspecting other possible pairing symmetries. Therefore, we propose a possible method to distinguish the pairing symmetry of the unconventional superconductors by means of the inelastic neutron scattering measurements.

It was argued that the long-range antiferromagnetic order found near the optimal electron-doped cuprates may arise from the impurities\cite{Motoyama-Nature07}. However, the AFM correlation length in the electron-doped cuprates is much longer than that in the hole-doped cuprates. Therefore, the AFM correlations plays important roles in the optimal electron-doped cup rates. Based on these considerations, we adopt an experimentally proposed SDW description to investigate the electron-doped cuprates near the optimal doping (For more details, See Ref.\cite{HY}). In the normal state, it can be simply expressed as
\begin{eqnarray}
H&=&\sum_{k\sigma }(\epsilon _{k}^{\prime }-\mu)(d_{k\sigma }^{+}d_{k\sigma
}+e_{k\sigma }^{+}e_{k\sigma }) +\sum_{k\sigma }\epsilon _{k}( d_{k\sigma
}^{+}e_{k\sigma }+h.c)  \nonumber \\
&&-\sum_{k\sigma }\sigma V_{\pi ,\pi }( d_{k\sigma }^{+}d_{k\sigma
}-e_{k\sigma }^{+}e_{k\sigma }).
\end{eqnarray}
In presence of SDW, the two sublattices $D$ and $E$ with respective fermionic operator $d$ and $e$ are introduced. $\epsilon _{k}=-2t(cosk_{x}+cosk_{y})$ and $\epsilon_{k}^{\prime}=-4t^{\prime}cosk_{x}cosk_{y}-2t^{\prime\prime}(cos2k_{x}+cos2k_{y})$ are inter- and intra-lattice hopping term with $t$, $t^{\prime}$, and $t^{\prime\prime}$ the nearest-neighbor (NN), second-NN, and third-NN hoping constant. $V_{\pi,\pi}$ is an effective $\mathbf{Q}$-scattering potential, representing the strength of the SDW. Its value can be phenomenologically evaluated by $V_{\pi,\pi}=UM$ with $U$ a reduced Coulomb repulsion and $M$ the antiferromagnetic order parameter in the mean-field level\cite{Das-PRB06,Luo-PRL05}. Here, we treat it as an independent parameter, which can be experimentally determined. The chemical potential $\mu$ is determined by the particle conservation.

In the SC state, a phenomenological BCS-like pairing
term $-\sum_{k}\Delta _{k}\left( d_{k\uparrow }e_{-k\downarrow
}+e_{k\uparrow}d_{-k\downarrow }+h.c.\right) $ with monotonic $d$-wave symmetry
$\Delta_{k}=\Delta(\cos k_{x}-\cos k_{y})$ is introduced. The quasiparticle dispersion is then
$E_{k}^{\eta}=\sqrt{(\xi_{k}^{\eta})^{2}+\Delta_{k}^{2}}$ with $\xi
_{k}^{\eta}=(\epsilon _{k}^{\prime}-\mu) +\eta \sqrt{\epsilon_{k}^{2}+V_{\pi ,\pi
}^{2}}$ ($\eta=1$, and $-1$ for upper, and lower band)\cite{Armitage-RMP}. The normal and anomalous Green's functions are both $2\times2$ matrices defined as $\hat{G}_{k\sigma}=-\langle T_{\tau} \psi_{k\sigma}(\tau)\psi_{k\sigma}^{\dagger}\rangle$, and $\hat{F}_{k}=-\langle T_{\tau}\psi_{-k\downarrow}(\tau)\psi_{k\uparrow}^{T}\rangle$, where $\psi_{k\sigma}=(d_{k\sigma}, e_{k\sigma})^{T}$.
The transversal spin susceptibility under the random phase approximation, also a $2\times2$ matrix, is expressed as
\begin{equation}
\hat{\chi}_{q}=\frac{\hat{\chi}_{q}^{0}}{1-U\hat{\chi}_{q}^{0}}
\end{equation}
with $U$ the above introduced reduced Coulomb repulsion. $\hat{\chi}_{q}^{0}=-\sum (\hat{G}_{k\downarrow}\hat{G}_{k+q\uparrow}+\hat{F}_{k}\hat{F}_{k+q}^{*})$ is the bare spin susceptibility, $k\equiv(k,\omega)$.

In numerics, the hoping constants are set as $t=250meV$, $t^{\prime}=-50meV$, and $t^{\prime\prime}=20meV$\cite{Parker-PRB07}. The doping density is set as $x=0.15$, near the optimal doping. The temperature is fixed at $0.2meV$, and $2meV$ for SC state, and normal state, respectively. We adopt the broaden factor $\Gamma=1meV$, comparable to the instrumental resolution.

First, it should be pointed out that the main features of magnetic excitations in the superconducting state, including the magnetic resonance $E^{M}_{res}$ and commensurability, keep almost unchanged in comparison with that in the normal state\cite{HY}. In order
to extract the role of superconductivity on magnetic excitations, we adopt the difference of spin
susceptibility near $\mathbf{Q}$ between the superconducting and normal state as $I_{\mathbf{Q}}(\omega)=\int_{\Omega }\left[
\Im \chi _{q}^{SC}(\omega )-\Im \chi _{\mathbf{q}}^{NM}(\omega )\right] dq$,
consisting with the experimental measurements\cite{Zhao-PRL07,Wilson-Nature06,Wilson-PNAS07}. The integration is restricted within a small region of $\pi/64 \times \pi/64$ centered around $\mathbf{Q}$ point. The main features are insensitive to the selected integral region.

\begin{figure}[btp]
\vspace{-0.0in} \hspace{-0.0in} \includegraphics[width=3.6in]{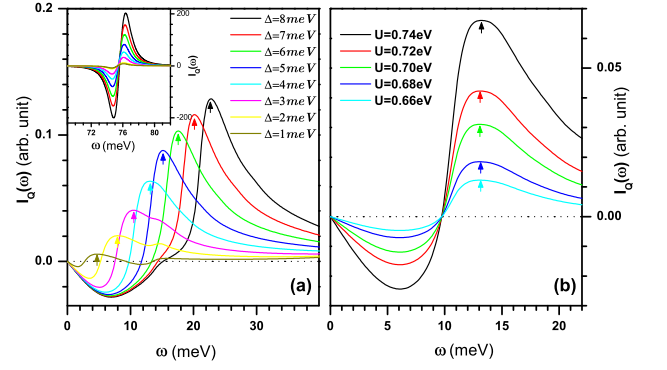}
\caption{(Color online) (a) Low energy dependence of $I_{\mathbf{Q}}(\omega)$ for different superconducting gap with $V_{\pi,\pi}=100meV$ and $U=0.72eV$. Insert in (a) is the high-energy dependence near the magnetic resonance. (b) Energy dependence of $I_{\mathbf{Q}}(\omega)$ for different $U$ with fixed superconducting gap $\Delta=4meV$. The intensity in $U=0.72eV$, $U=0.74eV$ had been re-scaled by a factor $1.5$, and $2.0$ for comparison, respectively. The superconducting resonance are denoted by arrows.}
\label{f.1}
\end{figure}

\begin{figure}[tbp]
\centering\includegraphics[width=3.6in]{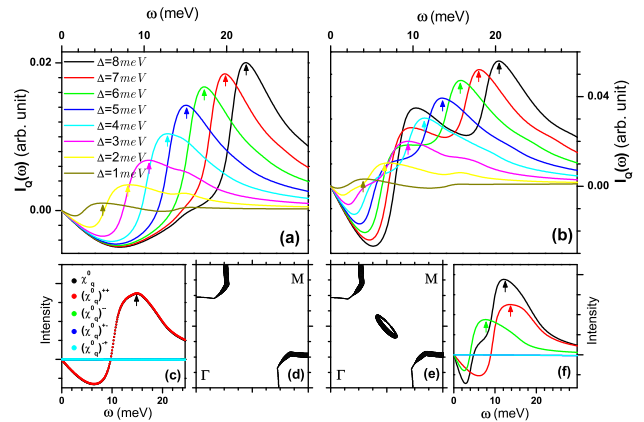}
\caption{(Color online) Upper panels: $I_{\mathbf{Q}}(\omega)$ as functions of energy for different superconducting gap with parameters $V=80meV$ in (a) and $V=50meV$ in (b), $U=0.6eV$. Arrows denote the resonance energy. Low panels: (c), and (f) are the bare spin susceptibility together with its four components with $\Delta=4meV$ as described in the text; (d), and (e) are the respective Fermi surface.}
\label{f.2}
\end{figure}

The typical $I_{\mathbf{Q}}(\omega)$ is shown in Fig.~\ref{f.1}(a). This choice is analogous to the situation that the oxygen is not well annealed as we shown before, where only the electron pocket near $(\pi, 0)$ can be found due to strong SDW (Fig.~\ref{f.2}(d)). $I_{\mathbf{Q}}(\omega)$ is negative at low enough energy region, which is referred as the spin gap region. It then increases gradually upon the energy and reaches its maximum at a given energy, where the superconducting resonance $E_{res}$ is experimentally defined. These low-energy features are qualitatively
consistent with the INS measurements on $NCCO$\cite{Zhao-PRL07} and $PLCCO$%
\cite{Wilson-Nature06,Wilson-PNAS07}. At higher energy region, a sharp dip near the magnetic resonance and then a peak can be found due to suppression of spin dynamics in the superconducting state. We focus on the low energy properties of the magnetic excitations below, which is closely related to the SC.

\begin{figure}[tbp]
\centering\includegraphics[width=3.6in]{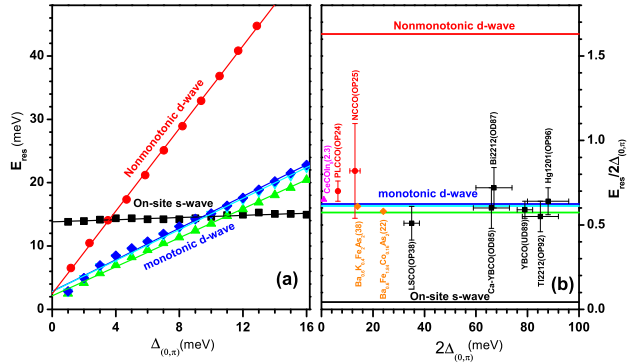}
\caption{(Color online) (a) Resonance energy $E_{res}$ as functions of superconducting gap $\Delta_{(0,\pi)}$. Red, and black symbols are for nonmonotonic $d$-wave, and on-site $s$-wave with selected parameters $U=0.72eV$ and $V_{\pi,\pi}=100meV$. Blue, cyan, and green symbols are for the monotonic $d$-wave, with respective effective $\mathbf{Q}$-scattering $V_{\pi,\pi}=100meV$, $80meV$, and $50meV$. $U=0.6eV$ for all. The solid lines are the linear fitting of respective data. (b) $E_{res}/2\Delta_{(0,\pi)}$ as functions of $2\Delta_{(0,\pi)}$. The solid lines are obtained from (a). All experimental data are directly extracted from Ref.~\cite{Yu-NPHY09}. The values in the parentheses are the superconducting critical temperature.}
\label{f.4}
\end{figure}

The resonance energy increases with the increasing superconducting gap $\Delta$. It is approximate to $9.5meV$,
and $12.5meV$ for the superconducting gap $\Delta=3meV$, and $4meV$. This is comparable to the experimental measurements in $PLCCO$\cite{Wilson-Nature06}. In fact, The superconducting resonance energy exhibits well linear
dependence on $\Delta$. The ratio $E_{res}/\Delta(0,\pi)$ is about $1.2$ (Fig.~\ref{f.4}), quantitatively consisting with the INS data\cite{Yu-NPHY09}. For strong enough superconductivity, the ratio may deviates from the linear dependence due to the overlap with magnetic resonance near the antiferromagnetic instability (not shown). Since $\frac{2\Delta}{k_{B}T_{c}}$ ranges from $3.5$\cite{Shan-PRB08,Dagan-PRB07} by the point contact and S-I-S planar tunneling measurement to $5.0$\cite{Homes-PRB06} by the optical properties measurements for optimal doping, the experimental linear scale of $E_{res}/T_{c} \sim 5.8$ is not hard to understood\cite{Wilson-Nature06}.

The superconducting resonance is insensitive to the selected parameters. To address it, different values of $U$ are adopted (Fig.~\ref{f.1}(b)). The low-energy magnetic excitations change from the commensurability to the incommensurability in the normal state\cite{HY}. The resonance energy $E_{res}$ is indeed unchanged for various reduced Coulomb repulsion. When the strength of SDW is suppressed, for example $V_{\pi,\pi}=80meV$ (Fig.~\ref{f.2}(a)), similar behaviors can be also found. The most important is that the linear scaling keeps almost unchanged as shown in Fig.~\ref{f.4} (a). Therefore, the superconducting resonance reflects the intrinsic nature of superconductivity. The ratio is also expected to be measured in underdoped and overdoped electron-doped cuprates. We also notice that the intensity of $I_{\mathbf{Q}}(\omega)$ weakens with decreasing $\Delta$. This is well consistent with INS measurements on $PLCCO$, where the external magnetic field is applied to suppress the superconductivity\cite{Wilson-PNAS07}.

Interestingly, when the strength of effective \textbf{Q}-scattering is reduced down to $50meV$, a weak but visible peak below $E_{res}$ emerges, especially for stronger SC (Fig.~\ref{f.2}(b)). Compared with $V_{\pi,\pi}=100meV$, the underlying Fermi surface develops from the electron pocket into large three-piece structure, i.e., the hole pocket is also present now\cite{Armitage-PRL01}. It is well known that the hole pocket develops when SDW is suppressed in the electron-doped cuprates. Meanwhile, the superconducting enhances. Therefore, this weak peak originates from the superconductivity near the hole pocket. To further confirm it, we decompose the bare spin susceptibility into four components, represented by the intra- ($(\chi_{Q}^{0})_{++}$ and $(\chi_{Q}^{0})_{--}$), and inter-band ($(\chi_{Q}^{0})_{+-}$ and $(\chi_{Q}^{0})_{-+}$) components. Obviously, the low energy peak comes from the contribution of lower-band (hole band). In comparison, the superconducting resonance comes from the upper-band (electron band). Recently, Yu \emph{et al.} found similar two-peaks structure in INS measurements on $NCCO$\cite{Yu-PRB10}. However, the low-, and high-energy peaks are thought to be associated the $A_{1g}$ and $B_{1g}/B_{2g}$ features in electronic Raman scattering\cite{Qazilbash-PRB05}, respectively. Here, the low-, and high-energy resonance is related to the $B_{2g}$, and $B_{1g}$ features. This suggests that the high-energy resonance may be separated into two sub-resonance, and the higher one follows the above mentioned linear scaling on $T_{c}$. We believe the present results can be discovered in the well oxygen annealed or slightly overdoped $n-$type cuprates. 

\begin{figure}[tbp]
\centering\includegraphics[width=3.6in]{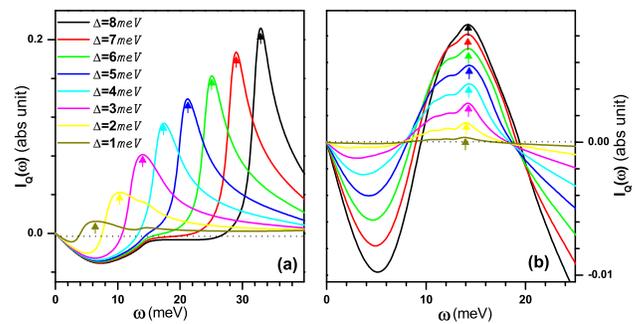}
\caption{(Color online) Energy dependence of $I_{\mathbf{Q}}(\omega)$ for different superconducting gap at given pairing symmetries. (a) is the nonmonotonic d-wave with third harmonic term, while (b) is the on-site s-wave. The parameters are selected as $U=0.72eV$ and $V=100meV$. Arrows denoting the peak position.}
\label{f.3}
\end{figure}

To understanding the relation between the resonance and superconductivity, we consider three other pairing symmetries. They are the nonmonotonic $d$-wave with third harmonic term $\Delta_{k}=\Delta(cosk_{x}-cosk_{y})-\Delta^{\prime}(cos3k_{x}-cos3k_{y})$ with $\Delta/\Delta^{\prime}=2.41$\cite{Kruger-PRB07}, extended s-wave $\Delta_{k}=\Delta (cosk_{x}+cosk_{y})$, and on-site s-wave $\Delta_{k}=\Delta$. The pairing in the latter occurs in the same lattice, whereas the former two occur between the different lattices, same as the monotonic $d$-wave. Similar well linear scales can be found in the nonmonotonic $d$-wave but with much enhanced ratio about $1.6$ (Fig.~\ref{f.3}(a) and Fig.~\ref{f.4}), much larger than the experimental data $0.6$\cite{Yu-NPHY09}. Therefore, the high harmonic term in $d$-wave superconductivity is not necessary when the SDW is considered as argued previously\cite{Luo-PRL05,Liu-PRB07}. For extended $s$-wave, no resonance can be found due to the vanishing of superconducting gap near the electron-pocket. For the on-site $s$-wave pairing symmetry (Fig.~\ref{f.3}(b) and Fig.~\ref{f.4}), the resonance is also present, but the resonance energy change little when the superconducting gap enhanced. Comparing with experimental data, the monotonic $d$-wave pairing is the most appropriate candidate. Therefore, the superconducting resonance is closely related to the pairing symmetry of superconducting gap, and then the mechanism of high-T$_{c}$ superconductors.

We notice that the resonance in the hole-doped cuprates and heavy fermion compounds\cite{Hegger-PRL00} follows the similar ratio. This manifests their pairing symmetry is also the monotonic $d$-wave, which had been confirmed previously. Interestingly, the Fe-pnictides share the same ratio. This fact challenges the so-called $s^{\pm}$-wave symmetry and expected to be studied further\cite{Ding-EPL08}. In this sense, we have proposed a feasible way to distinguish the pairing symmetry by INS measurements, which may be further applied on the other unconventional superconductors.

In conclusion, the superconducting resonance and its relation to superconductivity are studied in the optimal electron-doped cuprates within a SDW description. The main features in the superconducting state, discovered experimentally, are well established. The resonance energy exhibits well linear dependence on the superconducting gap, irrespective of the selected parameters. Therefore, the resonance is a universal feature of superconductivity. A sub-resonance peak, originating from the hole pockets, is further predicted when SDW is suppressed. The ratio of linear scaling is intimately related to the pairing symmetry. The value manifests the pairing symmetry in electron-doped cuprates is monotonic $d$-wave, same as that in hole-doped cuprates. Furthermore, we propose a possible method to distinguish the pairing symmetry of superconductivity by means of inelastic neutron scattering.

This work was supported by NSFC Projects No. 11274276, and A Project Funded by the Priority
Academic Program Development of Jiangsu Higher Education Institutions. CD Gong
acknowledges 973 Projects No. 2011CBA00102. HQ Lin acknowledges RGC Grant of HKSAR,
Project No. HKUST3/CRF/09.


\begin{thebibliography}{99}

\bibitem{Armitage-RMP} N. P. Armitage, P. Fournier, and R. L. Greene, Rev. Mod. Phys. \textbf{82}, 2421 (2010).

\bibitem{Zhao-NPHY11} J. Zhao \emph{et al.}, Nature Phys. \textbf{7}, 719 (2011).

\bibitem{Fujita-PRL08} M. Fujita \emph{et al.}, Phys. Rev. Lett. \textbf{101}, 107003 (2008).

\bibitem{Yamada-PRL03} K. Yamada \emph{et al.}, Phys. Rev. Lett. \textbf{90}, 137004 (2003).

\bibitem{Zhao-PRL07} J. Zhao \emph{et al.}, Phys. Rev. Lett. \textbf{99}, 017001 (2007).

\bibitem{Wilson-Nature06} S. D. Wilson \emph{et al.}, Nature \textbf{(}442), 59 (2006).

\bibitem{Wilson-PNAS07} S. D. Wilson \emph{et al.}, PNAS \textbf{104}, 15259 (2007).

\bibitem{Yu-NPHY09} G. Yu \emph{et al.}, Nature Phys. \textbf{5}, 873 (2009).

\bibitem{Yu-PRB10} G. Yu \emph{et al.}, Phys. Rev. B \textbf{82}, 172505 (2010).

\bibitem{FujitaRev} M. Fujita \emph{et al.}, J. Phys. Soc. Jpn. \textbf{81}, 011007 (2012).

\bibitem{Feng} L. Cheng and S. P. Feng, Phys. Rev. B \textbf{77}, 054518 (2008).

\bibitem{Ismer-PRL07} J.-P. Ismer \emph{et al.}, Phys. Rev. Lett, \textbf{99}, 047005 (2007).

\bibitem{LJX-PRB03} J. X. Li, J. Zhang, and J. Luo, Phys. Rev. B \textbf{68}%
, 224503 (2003).

\bibitem{Kruger-PRB07} F. Kr\"{u}ger \emph{et al.}, Phys. Rev. B \textbf{76%
}, 094506 (2007).

\bibitem{HY} H. Y. Zhang \emph{et al.}, arXiv:1212.1028.

\bibitem{Motoyama-Nature07} E. M. Motoyama \emph{et al.}, Nature \textbf{445}, 186 (2007).

\bibitem{Das-PRB06} T. Das, R. S. Markiewicz, and A. Bansil, Phys. Rev. B \textbf{74}, 020506(R) (2006).

\bibitem{Luo-PRL05} H. G. Luo and T. Xiang, Phys. Rev. Lett. \textbf{94}, 027001 (2005).

\bibitem{Parker-PRB07} S. R. Park \emph{et al.}, Phys.Rev.B, \textbf{75}, 060501(R) (2007).

\bibitem{Shan-PRB08} L. Shan \emph{et al.}, Phys. Rev. B \textbf{78}, 014505 (2008).

\bibitem{Dagan-PRB07} Y. Dagan and R. L. Greene, Phys. Rev. B \textbf{76}, 024506 (2007).

\bibitem{Homes-PRB06} C. C. Homes \emph{et al.}, Phys. Rev. B \textbf{74}, 214515 (2006).

\bibitem{Armitage-PRL01} N. P. Armitage \emph{et al.}, Phys. Rev. Lett. \textbf{87}, 147003 (2001).

\bibitem{Qazilbash-PRB05} M. M. Qazilbash \emph{et al.}, Phys. Rev. B \textbf{72}, 214510 (2005).

\bibitem{Liu-PRB07} C. S. Liu and W. C. Wu, Phys. Rev. B \textbf{76}, 014513 (2007).

\bibitem{Hegger-PRL00} H. Hegger \emph{et al.}, Phys. Rev. Lett. \textbf{84}, 4986 (2000).

\bibitem{Ding-EPL08} H. Ding \emph{et al.}, Europhys. Lett. \textbf{83}, 47001 (2008).

\end{thebibliography}
\end{document}